\documentclass[aps,prb,showpacs,amsmath,amssymb,twocolumn,superscriptaddress]{revtex4-1}

\usepackage{color}
\usepackage{amsmath}
\usepackage{graphicx}
\usepackage{multirow}

\begin{document}

\newcommand{\bn}{{\bf n}}
\newcommand{\bp}{{\bf p}}   
\newcommand{\br}{{\bf r}}
\newcommand{\bk}{{\bf k}}
\newcommand{\bv}{{\bf v}}
\newcommand{\brho}{{\bm{\rho}}}
\newcommand{\bj}{{\bf j}}
\newcommand{\wk}{\omega_{\bf k}}
\newcommand{\nk}{n_{\bf k}}
\newcommand{\eps}{\varepsilon}
\newcommand{\la}{\langle}
\newcommand{\ra}{\rangle}
\newcommand{\be}{\begin{eqnarray}}
\newcommand{\ee}{\end{eqnarray}}
\newcommand{\intl}{\int\limits_{-\infty}^{\infty}}
\newcommand{\dE}{\delta{\cal E}^{ext}}
\newcommand{\SE}{S_{\cal E}^{ext}}
\newcommand{\dsp}{\displaystyle}
\newcommand{\phit}{\varphi_{\tau}}
\newcommand{\p}{\varphi}
\newcommand{\cL}{{\cal L}}
\newcommand{\dphi}{\delta\varphi}
\newcommand{\dbj}{\delta{\bf j}}
\newcommand{\lra}{\leftrightarrow}

\newcommand{\rred}[1]{{#1}}
\newcommand{\skp}[1]{{#1}}
\newcommand{\bblue}{}

\title{Conductance of Interacting Quasi-One-Dimensional Electron Gas with a Scatterer}

\author{A.~V.~Borin}
\affiliation{Kotelnikov Institute of Radioengineering and Electronics, Mokhovaya 11-7, Moscow, 125009 Russia}
\affiliation{Moscow Institute of Physics and Technology, Institutsky per. 9, Dolgoprudny, 141700 Russia}

\author{K.~E.~Nagaev}
\affiliation{Kotelnikov Institute of Radioengineering and Electronics, Mokhovaya 11-7, Moscow, 125009 Russia}
\affiliation{Moscow Institute of Physics and Technology, Institutsky per. 9, Dolgoprudny, 141700 Russia}

\date{\today}

\begin{abstract}
We calculate the conductance of a quantum wire with two occupied subbands in a presence of a scatterer taking
into account the interaction between electrons. We extend the renormalization-group equation for the 
scattering matrix of the obstacle to the case of intersubband interactions, find its fixed points, and
investigate their stability. Depending on the interaction parameters, the conductance may be equal to 0,
$e^2/h$, or $2e^2/h$ per spin projection. In some parameter ranges, two stable fixed points may coexist,
so the ultimate conductance depends on the properties of the bare scatterer. For spinful electrons, the 
conductance of the wire may nonmonotonically depend on the Fermi level and temperature.
\end{abstract}

\maketitle

\section{Introduction}

Transport properties of one-dimensional (1D) quantum conductors are of significant interest because
they represent strongly correlated systems even for a weak electron - electron interaction \cite{Solyom78}. 
Despite this, 
the conductance of a uniform 1D wire smoothly connected to the leads is not changed by the interaction and 
equals $2e^2/h$ \cite{Maslov95,Safi95}. Things are different if
the wire contains an inhomogeneity that can backscatter electrons. In the case of repulsive
interaction, even a weak backscattering results in a power-law temperature dependence of conductance
and its complete suppression at zero temperature \cite{Kane92, Furusaki93}. Qualitatively, this behaviour may be explained as follows. 
The scattering off the inhomogeneity results in
Friedel oscillations of electron density with a period
$(2k_F)^{-1}$ and the corresponding oscillations of potential energy of electrons in the wire. 
As the interaction with potential oscillations couples the electron states with $+k_F$ and $-k_F$, it
would lead to a formation of the gap at the Fermi level and a completely insulating state if the
oscillations were of constant amplitude. However the amplitude falls off inversely proportionally to the distance
from the inhomogeneity, and this results only in a power-law suppression of the
conductance \cite{Matveev93} in long quantum wires at low temperatures. This suppression  was
observed in Refs. \onlinecite{Tarucha95} and \onlinecite{Chang96}. On the other hand, the attractive interaction
must result in a perfect transmission of the barrier \cite{Kane92}.

The properties of quantum wires with two populated transverse-quantization levels are even more complicated 
because of interaction between electrons in the different subbands. There are theoretical predictions 
\cite{Starykh00,Meyer07,Meyer09} that
these electron systems may exhibit zigzag ordering similar to a Wigner crystal and several gapped
phases may exist in them under different conditions. 
These phases result in different low-temperature behaviour of conductance of a wire with an impurity
\cite{Starykh00,Carr11}. 

The above theoretical considerations are based on the Luttinger-liquid model 
\cite{Giamarchi-b}, which allows one to 
exactly take into account the electron-electron forward scattering in the homogeneous wire. The elementary
excitations in this model are of bosonic  rather than fermionic type (plasmons and spinons).  However 
the electron scattering off an inhomogeneity cannot be treated exactly in this model. In addition,
it assumes a perfectly linear electron spectrum and cannot be used if the distance from the Fermi level
to the bottom of the subband is comparable with the interaction energy. The goal of this paper is to
use an approach similar to Refs. \onlinecite{Yue94,Lal02,Polyakov03,Titov06} and consider the scattering
off an inhomogeneity in a two-subband wire in the fermionic representation. We start 
from the exact scattering states of noninteracting electrons in a wire with the impurity and derive the
renormalization-group (RG) equations for its scattering matrix in the real space to one-loop order by assuming 
the interaction to be
weak. Then we find the stationary points of these equations and investigate their stability. In the most general
case, the interaction is described by three parameters, which correspond to the coupling in each of the 
two subbands and the intersubband coupling. For any set of parameter values, there is at least one stable
stationary point. Depending on these values, it corresponds to 0, 1, or 2  conductance quanta, so each of 
the two conducting channels is either fully open or fully closed at this point. In certain ranges of parameters,
two different stable fixed points may coexist. This indicates that the ultimate conductance depends on the
scattering matrix of the bare impurity. If one of the channels tends to become fully transmitting, and the 
other - fully reflecting, a nonmonotonic temperature dependence of the conductance may take place. Our results
differ from those of Refs. \onlinecite{Starykh00,Carr11}.


The paper is organized as follows. In Section II, we present the derivation of RG equations for the scattering 
matrix of the inhomogeneity for arbitrary number of conducting channel and arbitrary parameters of intra- and
interchannel interaction.
In Section III, we consider the particular case of a quantum wire with two conducting channels and find the 
fixed points of the RG flow. We investigate the behaviour of this flow near the fixed points and draw the diagram 
of their stability in the interaction-parameter plane. In Section IV, we consider the interaction parameters
for realistic systems with spinful electrons. In Section V, the temperature dependence of the conductance
is discussed, and Section VI presents the conclusion.

\section{RG equations}

Consider a long wire of uniform cross-section with a local inhomogeneity. The wire has $N$ transverse quantum
channels corresponding to transverse energies $E_i$ and transverse wave functions $\phi_i(y)$, so that the
total energy of an electron is the sum of transverse and longitudinal energies $E = E_i +\hbar^2 k_i^2/2m$.
The inhomogeneity is located at $x=0$ and described by the scattering matrix $S$, which relates the incoming 
and outgoing waves in different quantum channels. This matrix is unitary because of current conservation,
and we assume both time-reversal symmetry and a symmetric obstacle, so
\be 
 S = \begin{pmatrix}
         r & t \\
         t & r
     \end{pmatrix},
 \label{S}
\ee
where the $r$ and $t$ are symmetric $N \times N$ matrices of reflection and transmission amplitudes.
The scattering state with the electron incident in channel $i$ from the left is described by the wave
function
\be
 \Psi_{i}^{L}(\br,E) = \sum_{j} \frac{\Theta(E-E_j)}{\sqrt{2\pi\hbar v_j}}\,\psi_{ij}^{L}(x,E)\,\phi_{j}(y),
 \label{Psi}
\ee
where
\be 
 \psi_{ij}^{L}(x,E) = 
 \begin{cases}
    \delta_{ij}\,e^{ik_{i} x} + r_{ji}\,e^{-ik_{j} x}, & x<0
    \\
    t_{ji}\,e^{ik_j x}, & x>0
 \end{cases}.
 \label{psi}
\ee
and $v_j = \hbar^{-1}\,\partial E/\partial k_j$.
The scattering state with the electron incident from the right 
$\Psi_{i}^{R}(\br,E)$ is given by a similar expression with 
$\psi_{ij}^{R}(x,E) = \psi_{ij}^{L}(-x,E)$. These states obey the normalization condition
\begin{align}
 \int d\br\,\Psi_m^{\alpha *}(\br,E)\,\Psi_n^{\beta}(\br,E')
 =
 \delta_{\alpha\beta}\,\delta_{mn}\,\delta(E-E')
 \label{norm}
\end{align}
with $\alpha,\beta=L,R$ and form a full orthogonal basis for single-particle wave functions.

A weak electron-electron interaction results in a scattering of electrons by the Friedel oscillations caused by the inhomogeneity. To take it into account, we solve a Schr\"odinger-type equation
\be
\left[-\frac{\hbar^2}{2m}\nabla^2 + V_{eff}(\br)\right]\Psi(\br) = E\, \psi(\br)
\label{Schrodinger}
\ee
and obtain $\delta \Psi(\mathbf{r})$ in the lowest order in the interaction.

The interaction potential induced by the Friedel oscillations is a sum of a direct and exchange terms 
$V_{eff}(\br) = V_H(\br) - V_{ex}(\br)$, where
\begin{align}
V_{H} (\br) &= g_s \int{d\br_1 \, V(\br - \br_1) \, n(\br_1, \br_1)},
\label{V_H = U n()}
\\
V_{ex} (\br) \, \psi(\br) &= \int{d\br_1 \, V(\br - \br_1) \, n(\br, \br_1) \, \psi(\br_1)},
\label{V_ex = U n()}
\end{align}
$
 n({\bf r}, {\bf r_1}) = \langle \hat\Psi^{+}(\br_1)\,\hat\Psi(\br)\rangle
$
is the electron density matrix, and $V(\br - \br_1)$ is the potential of the electron-electron interaction. Typically, it is the Coulomb interaction screened by the gate.
The coefficient of spin degeneracy $g_s$ appears only in the direct term because it involves interactions between electrons with both spin directions while the exchange interaction is possible only for electrons with the same spin projection.

If the electron--electron interaction is considered as a perturbation, Eqs. (\ref{V_H = U n()}) and (\ref{V_ex = U n()}) result in a correction to the wave function (\ref{Psi}) in 
the form $\delta \Psi_i^{\alpha} = \delta \Psi_{i,H}^{\alpha} - \delta \Psi_{i,ex}^{\alpha}$, where 
\be 
 \delta\Psi_{i,H(ex)}^{\alpha}(\br) = \int d\br'\,G(\br,\br')\,V_{H(ex)}(\br')\,\Psi_i^{\alpha}(\br'),
 \label{dPsi}
\ee
and $G(\br,\br')$ is the Green's function of the single-particle Schr\"odinger equation.
The density matrix $n(\br,\br_1)$ is conveniently expressed in terms of the scattering states $\Psi_i^L$ and $\Psi_i^R$. At zero temperature, it is given by
\begin{equation}
 n(\br,\br_1) = \sum_{i,\alpha} \int_{-\infty}^{E_F} dE\,
    \Psi_i^{\alpha}(\br,E)\,[\Psi_i^{\alpha}(\br_1,E)]^{*}.
 \label{n}
\end{equation}
If $\Psi_i^{\alpha}$ are unperturbed by the interaction and given by Eqs. (\ref{Psi}) - (\ref{psi}),
the electron density far from the inhomogeneity may be presented in the form
\begin{align} 
 &n(\br,\br) = n_0 + \sum_{ij} \delta n_{ij}(x)\,\phi_i(y)\,\phi_j(y),
 \label{n(r)}\\
 \delta n_{ij}(x)&
 = |r_{ij}|
  \nonumber\\{}\times&
 \frac{\sqrt{v_{iF}\, v_{jF}}}{v_{iF} + v_{jF}}\,
 \frac{\sin[(k_{iF} + k_{jF})|x| + {\rm arg}\,r_{ij}]}{\pi |x|}.
 \label{n(x)}
\end{align}
The subscript $F$ shows that the corresponding quantity is evaluated at the Fermi level.
Apart from the uniform background $n_0$, $n(\br,\br)$ also contains rapidly oscillating components with wave
vectors $k_{iF} + k_{jF}$, which slowly decay with the distance from the inhomogeneity as $1/|x|$. The
presence of these components results in both interchannel and intrachannel backward scattering. 
Note that components with wave vectors $k_{iF} - k_{jF}$ vanish because of the unitarity condition
$r^{+}t + t^{+}r =0$.

The Green's function is conveniently expanded in transverse wave functions as
\be 
 G(\br, \br') = \sum_{i,j} g_{ij}(x,x')\,\phi_i(y)\,\phi_j(y'),
 \label{G}
\ee
where
\begin{multline}
 g_{ij}(x,x') = \frac{1}{i\hbar\sqrt{v_i v_j}}
 \\{}\times
 \begin{cases}
  t_{ij}\,e^{i|k_i x - k_j x'|}, & xx'<0
  \\ 
  \delta_{ij}\,e^{ik_i|x - x'|}   
  + r_{ij}\,e^{i|k_i x + k_j x'|}, & xx'>0
  \end{cases}
  \label{g}
\end{multline}
To obtain the corrections to the entries of the $S$ matrix, one has to substitute Eqs.
(\ref{n}) - (\ref{g}) into (\ref{dPsi}) and then to expand the resulting $\delta\Psi_i^{\alpha}$
in transverse wave functions $\phi_j(y)$. The presence of terms like (\ref{n(x)}) in the integrand
gives rise to terms that slowly decay as $1/|x|$ with the distance from the obstacle and a logarithmic
divergency of the integral at $E \to E_F$. Meanwhile it is precisely these values of energy that 
determine the conductance of the wire. To cure this divergency, we have to introduce a long-distance
cutoff $L_c$ and a short-distance cutoff $d$, which is of the order of the characteristic scale of the 
interaction potential $V(\br)$ (e.g of the distance from the conducting channel to the screening gate). 
As a result, one obtains the corrections to the reflection and transmission amplitudes in the form
\begin{align}
 &\delta r_{ij} = \frac{1}{2}\left[
  \alpha_{ij} r_{ij} - \sum_{kl}\alpha_{kl}\,( r_{ik} r_{kl}^{*} r_{lj} + t_{ik} r_{kl}^{*} t_{lj} )
 \right] \ln\frac{L_c}{d},
 \label{dr}\\
 &\delta t_{ij} = -\frac{1}{2}\sum_{kl} \alpha_{kl}\,
                 ( r_{ik} r_{kl}^{*} t_{lj} + t_{ik} r_{kl}^{*} r_{lj} )\,
                 \ln\frac{L_c}{d},
 \label{dt}
\end{align}
where different interactions are characterized by dimensionless parameters
\be 
\alpha_{ij}= (2-\delta_{ij})\,
\frac{V_{ijij}(0)-g_sV_{ijij}(k_{iF}+k_{jF})}{\pi \hbar (v_{iF}+v_{jF})},
\label{alpha}
\ee
which are expressed in terms of Fourier transforms of the interaction potential
\begin{multline}
 V_{klij}(x-x')=\int dy \int dy' \, \phi_{k}^*(y)\, \phi_{i}^*(y')
 \\{}\times
 V(x-x',y-y')\,\phi_{j}(y')\, \phi_{l}(y).
 \label{V}
\end{multline}
In what follows, we neglect the renormalization of the interaction parameters by backward scattering
obtained in Ref. \onlinecite{Solyom78} for spinful electrons because it plays no role in RG equations 
written to first  order in the interaction.

\begin{figure}[t]
 \includegraphics[width=8.5cm]{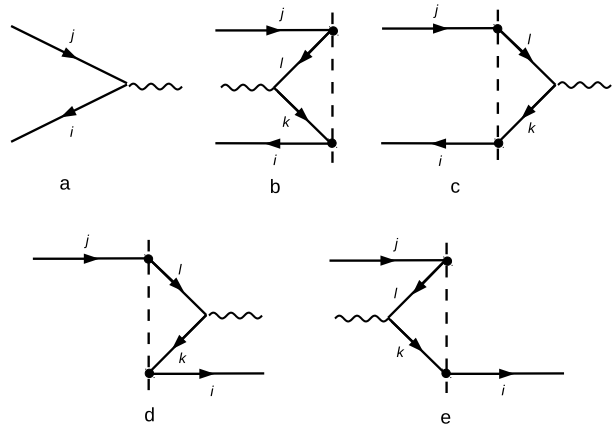}
 \caption{\label{fig:diagrams} Diagrams for the corrections to the reflection and transmission amplitudes.
 The vertices with wavy lines show scattering by Friedel oscillations, and the black dots show the scattering
 at the obstacle}
\end{figure}

The three terms in Eq. (\ref{dr}) are illustrated by the diagrams in Fig.~\ref{fig:diagrams}. The first term is presented 
by diagram $a$ and corresponds simply to backward scattering from channel $i$ to channel $j$ by Friedel oscillations.
The second term is presented by diagram $b$ and corresponds to reflection from state $j$ to state $l$ with 
subsequent backward scattering by Friedel oscillations from channel $l$ to channel $k$ and reflection from 
channel $k$ to channel $i$. The third term is presented by diagram $c$ and corresponds to a transmission
from channel $j$ to channel $k$, backscattering by Friedel oscillations from $l$ to $k$, and then transmission
from $k$ to $i$. Similar processes for $\delta t_{ij}$ are shown by diagrams $d$ and $e$.

As the first-order correction to the transmission and reflection amplitudes logarithmically diverges, one has
to sum up the perturbative series to infinite order. The summation of most divergent contributions may be 
performed using the renormalization procedure suggested by Yue et al. \cite{Yue94}. The idea is to choose the length $L_c$ in
such a way that the corrections $\delta r_{ij}$ and $\delta t_{ij}$ are still small and to treat the inhomogeneity together with Friedel oscillation within this distance as a new composite scatterer. 
Then the procedure is repeated again and again until  $S$ becomes independent of $L$.
By treating the ratio
$L/d$ as a continuous variable \cite{Yue94,Lal02} and using Eqs. (\ref{dr}) and (\ref{dt}), one obtains a differential equation for
the scattering matrix in the form
\be 
 \frac{dS}{dl} = SF^{+}S - F,
 \label{flow}
\ee
where $l=\ln(L/d)$ and the block-diagonal matrix
\be 
 F = \begin{pmatrix}
       f & 0\\
       0 & f
     \end{pmatrix}
 \label{F}
\ee
is formed of two identical $N \times N$ matrices with elements $f_{ij}= -\alpha_{ij} r_{ij}/2$. This equation
is similar to the RG equation obtained by Lal et al. \cite{Lal02} for a junction of several weakly
interacting wires, but $F$ contains now both diagonal and off-diagonal
elements. It is easily verified that the RG flow preserves the unitarity and symmetry of $S$. 

\begin{table*}[t]
\centering
\begin{tabular}{|c||c||c|c|c||c|c|c||c|c|c|c|c|c|}
\hline
$G/G_0$ & No & $|r_{11}|^2$ & $|r_{12}|^2$ & $|r_{22}|^2$ & $|t_{11}|^2$ & $|t_{12}|^2$ & $|t_{22}|^2$ 
 & $\lambda_1$ & $\lambda_2$ & $\lambda_3$ & $\lambda_4$ & $\lambda_5$ & $\lambda_6$
\\ \hline\hline\strut
\multirow{2}{*}{0} & I & 0 & 1 & 0 & 0 & 0 & 0 
 & $\frac{1}{2}(\alpha_{11}+\alpha_{22})-\alpha_{12}$
 & $\frac{1}{2}(\alpha_{11}+\alpha_{22})-\alpha_{12}$
 & $-\alpha_{12}$ & $-\alpha_{12}$ & $-\alpha_{12}$ 
 & $0\vphantom{\dfrac{1}{2}}$
\\ \cline{2-14}
 & II  & 1 & 0 & 1 & 0 & 0 & 0 
 & $-\frac{1}{2}(\alpha_{11}+\alpha_{22})$
 & $\alpha_{12}- \frac{1}{2}(\alpha_{11}+\alpha_{22})$
 & $-\alpha_{11}$ &  $-\alpha_{22}$ 
 & 0 & $0\vphantom{\dfrac{1}{2}}$
\\ \hline\hline\strut
\multirow{2}{*}{1} & III & 0 & 0 & 1 & 1 & 0 & 0 
 & $\frac{1}{2}(\alpha_{12}-\alpha_{22})$
 & $\frac{1}{2}(\alpha_{12}-\alpha_{22})$
 & $\alpha_{11}$ & $-\alpha_{22}$  
 & 0             & $0\vphantom{\dfrac{1}{2}}$
\\ \cline{2-14}
 & IV  & 1 & 0 & 0 & 0 & 0 & 1 
 & $\frac{1}{2}(\alpha_{12}-\alpha_{11})$
 & $\frac{1}{2}(\alpha_{12}-\alpha_{11})$
 & $-\alpha_{11}$ & $\alpha_{22}$  
 & 0 & $0\vphantom{\dfrac{1}{2}}$
\\ \hline\hline\strut
\multirow{2}{*}{2} & V & 0 & 0 & 0 & 0 & 1 & 0 
& $\frac{1}{2}(\alpha_{11}+\alpha_{22})$ & $\frac{1}{2}(\alpha_{11}+\alpha_{22})$ & $\alpha_{12}$ & 0
& 0 & $0\vphantom{\dfrac{1}{2}}$
\\ \cline{2-14}
& VI   & 0 & 0 & 0 & 1 & 0 & 1 
 & $\alpha_{11}$            & $\alpha_{22}$            & $\alpha_{12}$ & 0
 & 0 & $0\vphantom{\dfrac{1}{2}}$
\\ \hline
\end{tabular}
\caption{\label{table:isolated} Interaction-independent fixed points of RG. The number
 of conductance quanta per spin direction, the number of the fixed point, the scattering amplitudes, and the eigenvalues describing
 the behaviour of small deviations from these points.}
\end{table*}

In general, the symmetric $S$ matrix has $N^2+N$ different complex entries, which are reduced to the same 
number of independent real quantities because of the unitarity condition. Note that the form of Eq. (\ref{flow}) remains unchanged by the transformation 
\be
  S'=USU,
  \label{transform}
\ee
where $U$ is a diagonal $2N \times 2N$ matrix with elements $U_{ij} = \delta_{ij}\,\exp(i\p_j)$ and the real
phases $\p_j$ are independent of $l$ and obey the condition $\p_{j+N} = \p_j$. Therefore the matrices related
by Eq. (\ref{transform}) belong to the same family and evolve precisely in the same way. This suggests that 
the phases of $N$ elements of matrices $t$ and $r$ with different indices may be chosen arbitrarily when studying the RG flow.

\section{Fixed points}

Now we study the fixed points of RG flow Eq. (\ref{flow}) by considering the interaction parameters as formal quantities of arbitrary sign. The RG equation for the single-channel  wire with a scatterer has only two fixed points, which correspond to $r=1$ and $t=1$. The former is stable and the latter is 
unstable for the repulsive interaction while for the attractive one, they just exchange places.

In the case of two channels, the analysis is much more complicated because it involves six different entries
of $S$ matrix and three different interaction constants $\alpha_{11}$, $\alpha_{22}$, and $\alpha_{12}$. 
According to Eq. (\ref{flow}), the fixed points of RG should be determined from the condition
\be 
 SF^{+} = FS^{+},
 \label{fixed}
\ee
which results in four quadratic equations for $r_{ij}$ and $t_{ij}$. They should be supplemented by six
equations that follow from the unitarity of $S$ matrix. Fortunately, all the fixed points can be found 
analytically even for this case. It is easily shown (see Appendix) that all the three reflection amplitudes 
at the fixed points can be chosen as real numbers without a loss of generality. Any  fixed 
point can be obtained from one with real $r_{ij}$ by means of the transform (\ref{transform}), which involves
two independent phases $\p_1$ and $\p_2$.

Our analysis revealed three types of fixed points. First of all, there are four isolated universal points that
do not depend on the interaction parameters  and correspond to an integer number of conductance quanta
(points I - IV in Table \ref{table:isolated}). The second type includes several one-parametric families of fixed points that 
correspond to 
smooth transitions between the points of the first type (see Appendix, Table \ref{table:families}). One of these families
with $|r_{11}|=|r_{12}|=|r_{22}|=0$ and $|t_{11}|^2=|t_{22}|^2 =1-|t_{12}|^2$
corresponds to a transition between the point with $|t_{12}|=1$  and the point with $|t_{11}|=|t_{22}|=1$ 
(points V and VI in Table \ref{table:isolated}) and exists for any values of the interaction parameters. The rest 
of one-parameter families emerge only for definite
relations between $\alpha_{ij}$. Note the asymmetry between the transmission and reflection amplitudes. There is also a
third type of fixed points. These points are isolated, but the corresponding scattering amplitudes depend on 
the interaction parameters (see Appendix).

\begin{figure*}[t]
 \includegraphics[width=15cm]{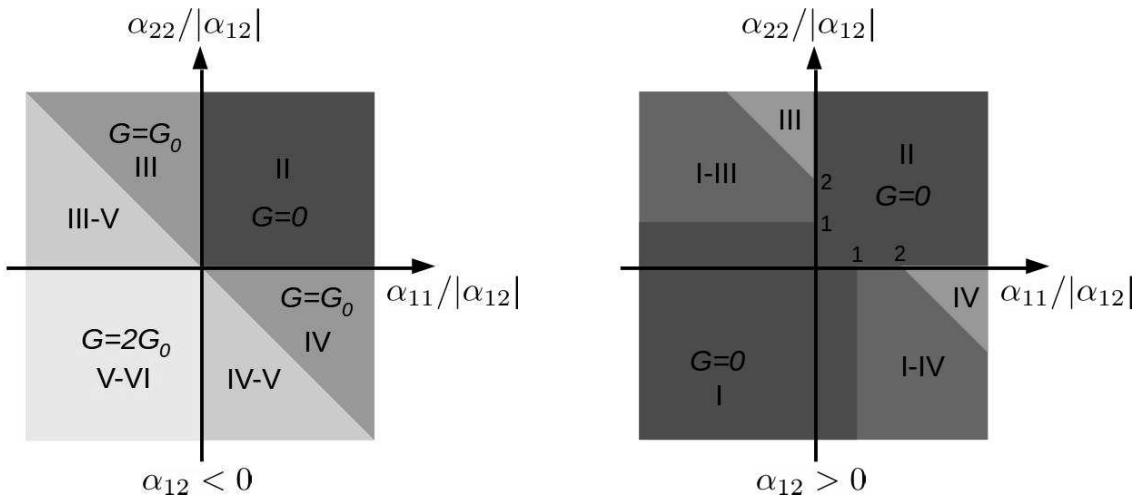}
 \caption{\label{fig:stability} The diagram of stability of isolated fixed points from 
 Table~\ref{table:isolated} for negative and positive $\alpha_{12}$ in the $\alpha_{11}$ - $\alpha_{22}$ plane.}
\end{figure*}

To investigate the stability of fixed points, we linearize Eq. (\ref{flow}) in $\delta r_{ij}$ and  
$\delta t_{ij}$ near them and eliminate the extra increments of these quantities using the unitarity 
condition. The evolution of the remaining six increments is determined by a $6\times 6$ matrix, and
the fixed point is stable if all the corresponding eigenvalues $\lambda_i$ are not positive. The eigenvalues describing the behaviour of small deviations of $S$ matrix from the isolated fixed 
points are listed in Table~\ref{table:isolated}. In general, 
this matrix has six eigenvalues, but some of them are zero because they correspond to the rotations of 
$S$ matrix (\ref{transform}), which involve two independent phases $\p_1$ and $\p_2$ and leave Eq. (\ref{flow})
invariant. As $S$ at points II - IV has two nonzero entries, there are two independent phase transformations 
and only four nonzero $\lambda$'s. At point I, $S$ has only one nonzero entry, so there is only one independent
phase rotation and five nonzero eigenvalues.  Points V and VI belong to the one-parametric continuous family of 
fixed points with zero reflectance and different $|t_{11}|=|t_{22}|$, and the three zero eigenvalues
at any point of this family correspond to variations of $|t_{ij}|$ and two phase rotations. The three nonzero eigenvalues 
for this family are given by equations 
\begin{subequations}
\label{lambda}
\begin{align}
 \lambda_{1}=&
 \frac{1}{2}\,
 (\alpha_{11} + \alpha_{22})\,(1-|t_{12}|)
 + \alpha_{12}\,|t_{12}|,
\\
 \lambda_{2,3}=&
 \frac{1}{4}
 \Bigl\{
   \alpha_{11}+\alpha_{22}+2\alpha_{12}
   +(\alpha_{11}+\alpha_{22}-2\alpha_{12})\,|t_{12}|
\nonumber\\&
   \pm
   \Bigl[\left(\alpha_{11}+\alpha_{22}\right)\left(1+|t_{12}|\right)+2\alpha_{12}\,(1-|t_{12}|)\Bigr]^2 
\nonumber\\& \qquad     
   -8\Bigl[ \alpha_{12}\,(\alpha_{11}+\alpha_{22})
\nonumber\\&
             +|t_{12}\,|(2\alpha_{11}\alpha_{22}-\alpha_{11}\alpha_{12}-\alpha_{22}\alpha_{12})
     \Bigr]^{1/2}
 \Bigr\}.
\end{align}
\end{subequations}
It is easily seen that fixed points of this family may be stable only if $\alpha_{12}<0$. If one goes
from larger to smaller values of $\alpha_{11}$ and $\alpha_{22}$, the first stable point with $|t_{12}|=1$
(point V in Table \ref{table:isolated}) appears at $\alpha_{11}+\alpha_{22}=0$. The range of stability 
broadens as these quantities decrease and finally includes the extreme point with $|t_{11}|=|t_{22}|=1$ 
(point VI in Table \ref{table:isolated}) as both the conditions $\alpha_{11}<0$ and $\alpha_{22}<0$ are 
fulfilled.

The regions of stability of the isolated fixed points from Table \ref{table:isolated} for different 
values of $\alpha_{ij}$ are shown in Fig.~\ref{fig:stability}. The darker areas 
correspond to lower conductance and the lighter areas, to higher conductance. It is easily seen that 
the diagram of stability
in the $\alpha_{11}$ - $\alpha_{22}$ plane essentially depends on the sign of $\alpha_{12}$. For 
$\alpha_{12}<0$, the conductance per spin projection varies from 0 to $2e^2/h$ as $\alpha_{11}$ and 
$\alpha_{22}$ decrease from $\infty$ to $-\infty$. If only one of these quantities tends to $-\infty$ and the 
other is positive,
two stable fixed points are possible. One of them is point III or IV with a conductance $e^2/h$, and 
the other is point V with a conductance $2e^2/h$. Hence the actual conductance, i.~e. the final point of RG 
flow  in these regions depends on where it starts, i.~e. 
on the scattering properties of the bare inhomogeneity.

 The diagram in $\alpha_{11}$ and $\alpha_{22}$
looks very different for $\alpha_{12}>0$. Surprisingly, the conductance is zero not only if $\alpha_{11}$ and 
$\alpha_{22}$ are both positive, but also if they are both negative. However it may equal $e^2/h$ if these 
quantities have opposite signs, so changing the sign of one of the interaction parameters from negative to positive may {\it increase} the conductance. 
There are also regions where stable points with $G=0$ and $G=e^2/h$ coexist. 

To conclude this section, the stationary points of RG for the two-subband wire are essentially the same as 
for two independent 1D
quantum wires, but the diagram of their stability is radically different from the latter case, especially for
positive $\alpha_{12}$.

\section{Interaction parameters for spinful electrons}

So far we considered the interaction parameters $\alpha_{ij}$ as arbitrary independent quantities. 
For actual physical interactions, this is not the case and the three parameters are related with each 
other. Consider a typical case of a quantum wire patterned in a two-dimensional electron gas by means 
of electrostatic gates. The Coulomb interaction between the electrons is screened by the gates and has the 
standard form
\be 
 V(r) = \frac{e^2}{\epsilon r} - \frac{e^2}{\epsilon\sqrt{r^2+4d^2}},
 \label{Coulomb}
\ee
where $\epsilon$ is the dielectric constant of the material and $d$ is the distance between the wire and
the gates, which determines the characteristic interaction length. It is evident that for spinless electrons 
with $g_s=1$,  all the three interaction parameters
 (\ref{alpha}) are positive because $V_{ijij}$ is a monotonically decreasing function of momentum transfer.
In this case, fixed point II with zero conductance is the only stable one.

\begin{figure}[t]
 \includegraphics[width=8.5cm]{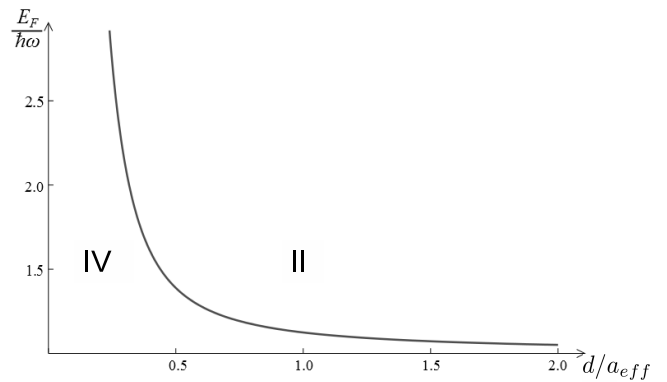}
 \caption{\label{fig:II-IV} The stability regions of fixed points II and IV in the $E_F$ - $d$ plane
 for the harmonic confining potential. The Fermi energy $E_F$ is normalized to the band separation
 $\hbar\omega$, and the distance to the gates $d$ is normalized to the effective width of the wire
 $a_{eff}=\sqrt{2\hbar/m\omega}$.}
\end{figure}

The case of spinful electrons with $g_s=2$ appears to be more interesting. If the Fermi level
crosses the upper subband not far from its bottom,  $k_{2F}$ may be much smaller than $1/d$, so
$V_{2222}(2k_{2F}) \approx V_{2222}(0)$ and $\alpha_{22}$ may be negative while $\alpha_{11}$  
is positive. This is just the case where fixed point IV is stable, so there is a full 
reflection of electrons in the lower subband and full transmission in the upper subband at zero temperature.
We determined the region of stability of this fixed point by calculating the interaction
parameters for a quantum wire with parabolic confinement and band separation $\omega$. The curve separating
the regions of stability of fixed points IV and II in the $E_F$ - $d$ plane obtained by numerical calculations
is shown in Fig.~\ref{fig:II-IV}. The Fermi energy is normalized to the
separation between the energy levels in the transverse harmonic potential $\hbar\omega$, and $d$ is
normalized to the effective width of the quantum wire $a_{eff}=\sqrt{2\hbar/m\omega}$. At small $d$, the 
separation curve follows the law $E_F/\hbar\omega \propto (a_{eff}/d)^2$.

The transition from fixed point IV to II may be observed by changing the Fermi level in the system by means 
of the gate voltage. If $\alpha_{11}$ is positive for $E_F \lesssim\omega$, the transmission of electrons in 
the lower subband is fully suppressed. Crossing the bottom of the upper subband by the Fermi level will bring
the wire to the fixed point IV with a conductance $2e^2/h$, and further increase of $E_F$ will result in 
crossing the separation line between points IV and II and returning to zero conductance. Therefore the 
electron - electron interaction may lead to a spike in the dependence of conductance on $E_F$.

\section{Temperature dependence of the conductance}

The interaction effects in two-channel quantum wires may manifest themselves as a nontrivial temperature 
dependence of the conductance. At a finite temperature the RG flow (\ref{flow}) must be stopped when the cutoff
length reaches the distance at which the Friedel oscillations are smeared out by thermal broadening
of the Fermi step. In the case of a multichannel wire, there is a set of lengths 
$L_{ij}^T = \hbar(v_{iF} + v_{jF})/T$ at which the interaction parameters $\alpha_{ij}$ turn into zero.
As $L$ increases, Eq. (\ref{flow}) must be solved separately
in each interval between the consecutive $L_{ij}^T$. The  interaction parameters corresponding to
smaller $L_{ij}^T$ must be set equal to zero and the value of $S$ in the end of previous interval must be used 
as the initial condition for the RG equation. The procedure must be stopped as $L$ reaches 
$L_{max}^T \equiv {\rm max}(L_{ij}^T)$, and $S(L_{max}^T)$ gives the conductance at temperature $T$.

\begin{figure}[t]
 \includegraphics[width=8cm]{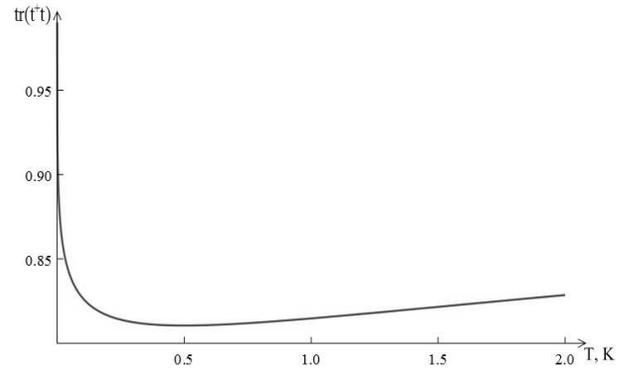}
 \caption{\label{fig:temperature} The temperature dependence of dimensionless conductance for a quantum
 wire with $\hbar\omega/E_F = 0.98$ and $d/a_{eff}=1.5$. The initial transmission amplitudes are 
 $r_{12}=t_{12}=0$, $t_{11}=0.65i$, $r_{22}=0.6i$, $r_{11}=\sqrt{1 - |t_{11}|^2}$, and
 $t_{22}=\sqrt{1 - |r_{22}|^2}$.
 }
\end{figure}

If the RG flow tends to point IV, the conductance of the lower subband tends to zero, and
the conductance of the upper subband tends to unity. In the vicinity of this point, the deviations from these
quantities decrease with increasing cutoff length $L$ according to power laws with exponents $\lambda_i$ from 
Table \ref{table:isolated}.  As the exponents for lower and upper subbands are different, a nonmonotonic 
temperature dependence of the conductance may be observed if the initial scattering amplitudes are chosen
appropriately. 

The results of numerical simulations for the parameters typical of GaAs heterostructures are 
shown in Fig.~\ref{fig:temperature}. The Fermi energy is $E_F=14$ meV, the subband separation is $\hbar\omega
=0.98\,E_F$, and the distance to the gate is $d=1.5\, a_{eff}$. A substitution of these values into Eqs. 
(\ref{V}) and (\ref{alpha}) gives the interaction parameters $\alpha_{11}=0.106$, $\alpha_{12}=-0.01$, and
$\alpha_{22}=-0.28$. Assume that the bare inhomogeneity is described by the scattering matrix with
$r_{12}=t_{12}=0$, $t_{11}=0.65i$, $r_{22}=0.6i$, $r_{11}=\sqrt{1 - |t_{11}|^2}$, and
 $t_{22}=\sqrt{1 - |r_{22}|^2}$.
A solution of Eq. (\ref{flow}) with these initial conditions leads to a maximum in the 
conductance at $T \sim 0.4$ K.

\section{Conclusion}

In summary, we have considered the effects of electron - electron interaction on the conductance of a 
quantum wire with an obstacle. The obstacle gives rise to Friedel oscillations of electron density, 
which cause backward scattering of electrons both within the transverse subbands and from one subband to another. 
In the case of two populated transverse subbands, the conductance of the wire is determined by two intraband
and one interband interaction parameter. Each subband is either in a fully insulating or fully conducting 
state as in 
the case of zero interaction between them, but which of these states will be realized depends on all the three
interaction parameters.  The wire has zero conductance  if all the three parameters are positive and maximum possible conductance $2e^2/h$ per spin direction if all of them are negative. However the conductance is zero
if the interband parameter is positive and both intraband parameters are negative. Changing the sign of one of 
them from negative to positive results in an increase in the conductance from 0 to $e^2/h$ per spin direction.

For spinful electrons with Coulomb interaction, the interaction parameters may change their signs depending
on the filling of the subbands. This may result in a nonmonotonic dependence of the conductance on the position
of the Fermi level in the wire. The temperature dependence of the conductance may be also nonmonotonic if the intraband interaction parameters have opposite signs.

Our results differ from those obtained in the Luttinger-liquid model because the limiting transitions are made
in a different sequence. In the Luttinger-liquid papers, the length of the wire was assumed to be infinite 
from the very beginning, and therefore the gaps in the spectrum developed even for a weak  interaction, though
they were exponentially small in this case. Nevertheless these gaps were essential because the low-energy limit
was considered  there. In contrast to this,  we perform calculations for a wire of a finite length in the limit
of weak interactions and only then let this length to infinity. This is why the gaps are washed out by a finite
dwell time of an electron in the wire. Similar results can be obtained by starting with an infinite wire if the temperature is assumed to be much higher than the gap width.

\begin{acknowledgments}
We are grateful to S. T. Carr and B. N. Narozhny for a useful discussion.
This work was supported by Russian Foundation for Basic Research, grant 13-02-01238-a, and by the program of Russian Academy of Sciences.
\end{acknowledgments}

\appendix

\section{\label{sec:details} Interaction-dependent fixed points of RG}

If the obstacle is symmetric and the number of channels is $N=2$, the scattering matrix has the general form
\be
 S = 
     \begin{pmatrix}
      r_{11} & r_{12} & t_{11} & t_{12} \\
      r_{12} & r_{22} & t_{12} & t_{22} \\
      t_{11} & t_{12} & r_{11} & r_{12} \\
      t_{12} & t_{22} & r_{12} & r_{22}
     \end{pmatrix}.
 \label{S_symm}
\ee
The unitarity condition $S S^{+} = 1$ leads to a system of six independent equations
\begin{subequations}\label{unit_eq}
\begin{align}
 &|r_{11}|^2 + |r_{12}|^2 + |t_{11}|^2 + |t_{12}|^2 = 1 \label{u1}\\
 &|r_{12}|^2 + |r_{22}|^2 + |t_{12}|^2 + |t_{22}|^2 = 1 \label{u2}\\
 &r_{11} r_{12}^{*} + r_{12} r_{22}^{*} + t_{11} t_{12}^{*} + t_{12} t_{22}^{*} = 0 \label{u3}\\
 &r_{11} t_{11}^{*} + r_{12} t_{12}^{*} + t_{11} r_{11}^{*} + t_{12} r_{12}^{*} = 0 \label{u4}\\
 &r_{11} t_{12}^{*} + r_{12} t_{22}^{*} + t_{11} r_{12}^{*} + t_{12} r_{22}^{*} = 0 \label{u5}\\
 &r_{12} t_{12}^{*} + r_{22} t_{22}^{*} + t_{12} r_{12}^{*} + t_{22} r_{22}^{*} = 0.\label{u6}
\end{align}
\end{subequations}
Matrix $F$ in Eq. (\ref{flow}) is of the form 
\be 
 F = 
     \frac{1}{2}
     \begin{pmatrix}
      -\alpha_{11} r_{11} & -\alpha_{12} r_{12} &  0                 & 0
      \\
      -\alpha_{12} r_{12} & -\alpha_{22} r_{22} &  0                 & 0
      \\
      0                  & 0                  & -\alpha_{11} r_{11} & -\alpha_{12} r_{12}
      \\
      0                  & 0                  & -\alpha_{12} r_{12} & -\alpha_{22} r_{22}
     \end{pmatrix}.
 \label{F2}
\ee
Hence the condition Eq. (\ref{fixed}) for the existence of a fixed point leads to a system of 
another four 
independent equations
\begin{subequations}\label{stat_eq}
\begin{align}
 &\alpha_{11}\, r_{11} r_{12}^{*} + \alpha_{12}\, r_{12} r_{22}^{*}
 =\alpha_{12}\, r_{11} r_{12}^{*} + \alpha_{22}\, r_{12} r_{22}^{*}
 \label{s1}\\
 &\alpha_{11}\, r_{11} t_{11}^{*} + \alpha_{12}\, r_{12} t_{12}^{*}
 =\alpha_{11}\, t_{11} r_{11}^{*} + \alpha_{12}\, t_{12} r_{12}^{*}
 \label{s2}\\
 &\alpha_{11}\, r_{11} t_{12}^{*} + \alpha_{12}\, r_{12} t_{22}^{*}
 =\alpha_{12}\, t_{11} r_{12}^{*} + \alpha_{22}\, t_{12} r_{22}^{*}
 \label{s3}\\
 &\alpha_{12}\, r_{12} t_{12}^{*} + \alpha_{22}\, r_{22} t_{22}^{*}
 =\alpha_{12}\, t_{12} r_{12}^{*} + \alpha_{22}\, t_{22} r_{22}^{*}
 \label{s4}
\end{align}
\end{subequations}
Equation (\ref{flow}) is invariant with respect to the transform (\ref{transform}), where
\be 
 U = 
     \begin{pmatrix}
      e^{i\p_1}          & 0                  &  0                 & 0
      \\
      0                  & e^{i\p_2}          &  0                 & 0
      \\
      0                  & 0                  &  e^{i\p_1}         & 0
      \\
      0                  & 0                  &  0                 & e^{i\p_2}
     \end{pmatrix}
 \label{U}
\ee
The reflection and transmission amplitudes are changed by this transform as follows:
\be
\begin{aligned}
 r_{11}' & = r_{11}\,e^{2i\p_1},     & \qquad  t_{11}' & = t_{11}\,e^{2i\p_1} \\
 r_{22}' & = r_{22}\,e^{2i\p_2},     & \qquad  t_{22}' & = t_{22}\,e^{2i\p_2} \\
 r_{12}' & = r_{12}\,e^{i\p_1+i\p_2},& \qquad  t_{12}' & = t_{12}\,e^{i\p_1+i\p_2}.
\end{aligned}
\label{transform2}
\ee
This allows us to make $r_{11}$ and $r_{22}$ real by appropriately choosing
$\p_1$ and $\p_2$. Therefore Eq. (\ref{s1}) is brought to the form
\begin{equation}
 (\alpha_{11} - \alpha_{12})\,r_{11}\,r^{*}_{12}
 +
 (\alpha_{12} - \alpha_{22})\,r_{22}\,r_{12} = 0,
 \label{s1a}
\end{equation}
which suggests that $r_{12}$ is either purely real or purely imaginary. We choose it
to be purely real because in the opposite case, it can be made such just by adding $\pi$ either to 
$\p_1$ or $\p_2$ in the transform (\ref{transform}).

\begin{table*}[t]
\centering
\begin{tabular}{|c|c||c|c|c||c|c|c|}
\hline
Path & Condition & $|r_{11}|^2$ & $|r_{12}|^2$ & $|r_{22}|^2$ & $|t_{11}|^2$ & $|t_{12}|^2$ & $|t_{22}|^2$ 
\\ \hline\hline\strut
$5\lra 6$ &  none            &
                   0 & 0 & 0 & $1-|t_{12}|^2$ & $0\div1$ & $1-|t_{12}|^2$ \\ \hline 
$2\lra 4$ & $\alpha_{11}=0$  &
          $0\div 1$  & 0 & 1 & $1-|r_{11}|^2$ & 0 & 0  \\ \hline
$3\lra 6$ & $\alpha_{11}=0$  &    
           $0\div 1$ & 0 & 0 & $1-|r_{11}|^2$ & 0 & 1  \\ \hline       
$1\lra 5$ &  $\alpha_{12}=0$ &
                   0 & $0\div 1$ & 0          & 0 & $1-|r_{12}|^2$ & 0\\ \hline 
$1\lra 6$ &  $\alpha_{12}=0$ &
                    0 & $0\div1$  & 0    & $1-|r_{12}|^2$ & 0 & $1-|r_{12}|^2$ \\ \hline                    
$2\lra 3$ &  $\alpha_{22}=0$ &
                    1 & 0 & $0\div1$ & 0 & 0  & $1-|r_{22}|^2$ \\ \hline
$4\lra 6$ &  $\alpha_{22}=0$ &   
                    0 & 0 & $0\div1$ & 1 & 0  & $1-|r_{22}|^2$ \\ \hline               
$2\lra 5$ &  $\alpha_{11}+\alpha_{22}=0$ &  
       $1-|t_{12}|^2$ & 0 & $1-|t_{12}|^2$ & 0 & $1\div0$ & 0 \\ \hline  
$1\lra 2$ &  $\alpha_{11}+\alpha_{22}=2\alpha_{12}$ &
                 $1-|r_{12}|^2$ & $0\div 1$ & $1-|r_{12}|^2$          & 0 & 0 & 0\\ \hline                             
\end{tabular}  
\caption{\label{table:families} One-parametric families of fixed points of RG. The first column indicates
the extreme points of the family according to Table \ref{table:isolated}, and the second column shows the
condition for the existence of the family.}
\end{table*}

If one or more of the scattering amplitudes is zero, Eqs. (\ref{unit_eq}) and (\ref{stat_eq}) are easily
solved, and the corresponding interaction-independent solutions are listed in Table \ref{table:isolated}. 
In addition to them, there are fixed points that exist only for the specific combinations of interaction 
parameters, which are not considered here.
Assume now that none of the scattering amplitudes are zero. Express $r_{22}$ in terms of $r_{11}$ by means of 
Eq. (\ref{s1a}), and the real and imaginary parts of $t_{12}$ and $t_{11}$ by means of Eqs. (\ref{u4}),
(\ref{u6}), (\ref{s2}), and (\ref{s4}). Substituting these quantities into the imaginary part of Eq. (\ref{u3})
results in the equation
\begin{multline}
 \left\{
   \frac{\alpha_{22}}{\alpha_{12}}
   \left[
     1 -\frac{\alpha_{11}-\alpha_{12}}{\alpha_{22}-\alpha_{12}}
   \right]
  -\left[
     1-\frac{\alpha_{22}(\alpha_{11}-\alpha_{12})}{\alpha_{11}(\alpha_{22}-\alpha_{12})}
    \right]
 \right\}
 \\ \times
 {\rm Im}\,t_{12}\,{\rm Re}\,t_{12}=0.
\label{either}
\end{multline}
Hence for arbitrary interaction parameters, $t_{22}$ has either zero real or imaginary part.
In the case of real $t_{22}$, the solution is
\begin{subequations}
\begin{gather}
 r_{12} =  \mp t_{12} = \pm\frac{\sqrt{(\alpha_{11}-\alpha_{12})(\alpha_{22}-\alpha_{12})}}
                             {\alpha_{11}+\alpha_{22}-2\alpha_{12}} {}\\
 r_{11} =  \pm t_{22} =     \frac{\alpha_{22}-\alpha_{12}}
                             {\alpha_{11}+\alpha_{22}-2\alpha_{12}} {}\\
 t_{11} =  \pm r_{22} =     \frac{\alpha_{11}-\alpha_{12}}
                             {\alpha_{11}+\alpha_{22}-2\alpha_{12}} {},
\end{gather}
\label{real}
\end{subequations}

If $t_{22}$ is imaginary, the scattering amplitudes are given by equations
\begin{subequations}
 \label{imaginary}
\be 
 r_{11}=\frac{\alpha_{12}\,(\alpha_{22}-\alpha_{12})}{\alpha_{12}^2-\alpha_{11}\alpha_{22}},
\ee
\be 
 r_{22}=\frac{\alpha_{12}\,(\alpha_{11}-\alpha_{12})}{\alpha_{12}^2-\alpha_{11}\alpha_{22}},
\ee
\begin{multline}
 r_{12}=(-1)^m 
 \\ \times
 \frac{\sqrt{\alpha_{11}\alpha_{22}\,(\alpha_{11}-\alpha_{12})(\alpha_{22}-\alpha_{12})}}   
                   {\alpha_{12}^2-\alpha_{11}\alpha_{22}},
\end{multline}
\begin{multline}
 t_{11}=i (-1)^n\frac{\alpha_{22}\,(\alpha_{11}-\alpha_{12})} 
                     {\alpha_{12}^2-\alpha_{11}\alpha_{22}}
\\    \times                  
    \sqrt{
       \frac{\alpha_{12}\,(\alpha_{22}+\alpha_{11}-2\alpha_{12})}
            {\alpha_{11}(\alpha_{22}-\alpha_{12})+\alpha_{22}(\alpha_{11}-\alpha_{12})}
         },
\end{multline}
\begin{multline}
 t_{22}=i (-1)^n  \frac{\alpha_{11}\,(\alpha_{22}-\alpha_{12})} 
                    {\alpha_{12}^2-\alpha_{11}\alpha_{22}}
\\  \times                  
 \sqrt{
       \frac{\alpha_{12}\,(\alpha_{22}+\alpha_{11}-2\alpha_{12})}
            {\alpha_{11}(\alpha_{22}-\alpha_{12})+\alpha_{22}(\alpha_{11}-\alpha_{12})}
      },
\end{multline}
\begin{multline}
 t_{12}=i (-1)^{n+m} \frac{\sqrt{\alpha_{11}\alpha_{22}(\alpha_{11}-\alpha_{12})(\alpha_{22}-\alpha_{12})}}  
                         {\alpha_{12}^2-\alpha_{11}\alpha_{22}}
\\  \times                       
 \sqrt{
      \frac{\alpha_{12}\,(\alpha_{22}+\alpha_{11}-2\alpha_{12})}
           {\alpha_{11}(\alpha_{22}-\alpha_{12})+\alpha_{22}(\alpha_{11}-\alpha_{12})}
      }, 
\end{multline}
\end{subequations}
where $n,m=1,2$. Fixed points (\ref{real}) and (\ref{imaginary}) are unstable for any interaction parameters.

\end{document}